\documentclass[aps,prl,twocolumn,floatfix,showpacs]{revtex4-1}
\usepackage{amsmath,amssymb,bm,epsfig,enumerate}
\usepackage{epsfig,xcolor,tikz}
\usepackage{bm} 

\renewcommand{\it}[1]{\textit{#1}}

\begin{document}
\title{Spinodal decomposition in homogeneous and isotropic turbulence}

\author{Prasad Perlekar$^{1,2}$, Roberto Benzi$^{3}$, Herman
  J.H. Clercx$^{1}$, David R. Nelson$^{4}$, and Federico
  Toschi$^{1,5}$}

\affiliation{$^{1}$ Department of Physics and Department of
  Mathematics and Computer Science and J.M. Burgerscentrum, Eindhoven
  University of Technology, 5600 MB Eindhoven, The Netherlands; and
  International Collaboration for Turbulence Research}
\affiliation{$^{2}$ TIFR Centre for Interdisciplinary Sciences, 21
  Brundavan Colony, Narsingi, Hyderabad 500075, India}
\affiliation{$^{3}$ Dip. di Fisica and INFN, Universit\`a ``Tor
  Vergata", Via della Ricerca Scientifica 1, I-00133 Roma, Italy}
\affiliation{$^{4}$ Lyman Laboratory of Physics, Harvard University,
  Cambridge, MA 02138, USA} \affiliation{$^{5}$ CNR, Istituto per le
  Applicazioni del Calcolo, Via dei Taurini 19, 00185 Rome, Italy}

\begin{abstract}
We study the competition between domain coarsening in a symmetric binary
mixtures below the critical temperature and turbulent fluctuations.  We
find that the coarsening process is arrested in presence of
turbulence. The physics of the process shares remarkable similarities
with the behaviour of diluted turbulent emulsions and the arrest
length scale can be estimated with an argument similar to the one
proposed by Kolmogorov and Hinze for the maximal stability diameter of
droplets in turbulence. Although in the absence of flow the microscopic 
diffusion constant is negative, turbulence does effectively arrest the inverse cascade of concentration fluctuations by making the low wavelength diffusion constant positive for scales above the Hinze length.
 \end{abstract}

\maketitle

Turbulence is known to strongly increase mixing efficiency. A scalar
concentration field in a turbulent flow undergoes a cascade of
stretching and folding processes that transfers concentration
gradients from large- to small-scales, resulting in efficient mixing
on times-scales much shorter that those associated with diffusion. The
enhanced mixing properties of turbulence arises due its multi-time and
multi-scale correlated velocity fluctuations and can be understood in
terms of a phenomenological, and scale-dependent, eddy-viscosity
$\nu_{t}(\ell)\sim \nu (\ell/\eta)^{4/3}$ ($\eta$ is the Kolmogorov
dissipative scale at which velocity fluctuations are dissipated
\cite{ric26}).  At larger inertial length scales, $\ell>\eta$, the
effective diffusivity $\nu_t\gg \nu$ where $\nu$ is the kinematic
viscosity of the quiescent fluid.

It is well known that a binary liquid mixture cooled below its
critical temperature undergoes a phase transition and the mixture
separates into phases enriched with its two components. This
phenomenon is known as spinodal decomposition.  Theoretically, the
temperature below which the system undergoes the phase transition is
determined by the point where the free-energy minima corresponding to
the two phases degenerates \cite{Dill03}. The dynamics of the phase
separation can be understood in terms of incompressible Navier-Stokes
equations coupled to a Cahn-Hilliard or model-B equations describing
the binary mixture order parameter (in the absence of
turbulence)~\cite{cah68, hal77}.  Using dimensional estimates, the
evolution of the phase separation can be divided into three regimes:
$(a)$ In the initial state, the coarsening length scale of the phase
separating binary mixture grows as $t^{1/3}$ (\it{Lifshitz-Slyozov
  scaling}, \cite{lif59}). This corresponds to growth dominated by the
binary mixture diffusivity and is associated with the evaporation of
small droplets at the expense of larger growing ones. $(b)$ At
intermediate times, when fluid motion becomes important, and viscous
dissipation of the fluid balances the pressure $(\nu\nabla^2 {\bm
  u}\sim \nabla p)$ which leads to a linear increase $\sim t$ in the
coarsening length (\it{Viscous scaling}, \cite{sig79}). $(c)$ At final
stages, the coarsening length scale grows as $t^{2/3}$ and is governed
by the balance of fluid advection with the variations in chemical
potential $(\rho {\bm u}\cdot \nabla {\bm u} \sim \nabla \mu)$
(\it{Inertial scaling}, \cite{fur85}).  This evolution of the
coarsening process has been verified in earlier numerical \cite{loo96,
  ken00, pag01, seg03} and experimental studies \cite{hob06}.

\begin{figure*}
    \centering
    \begin{tikzpicture}
    \node[above right] (img) at (0,0) {
\includegraphics[width=0.23\linewidth]{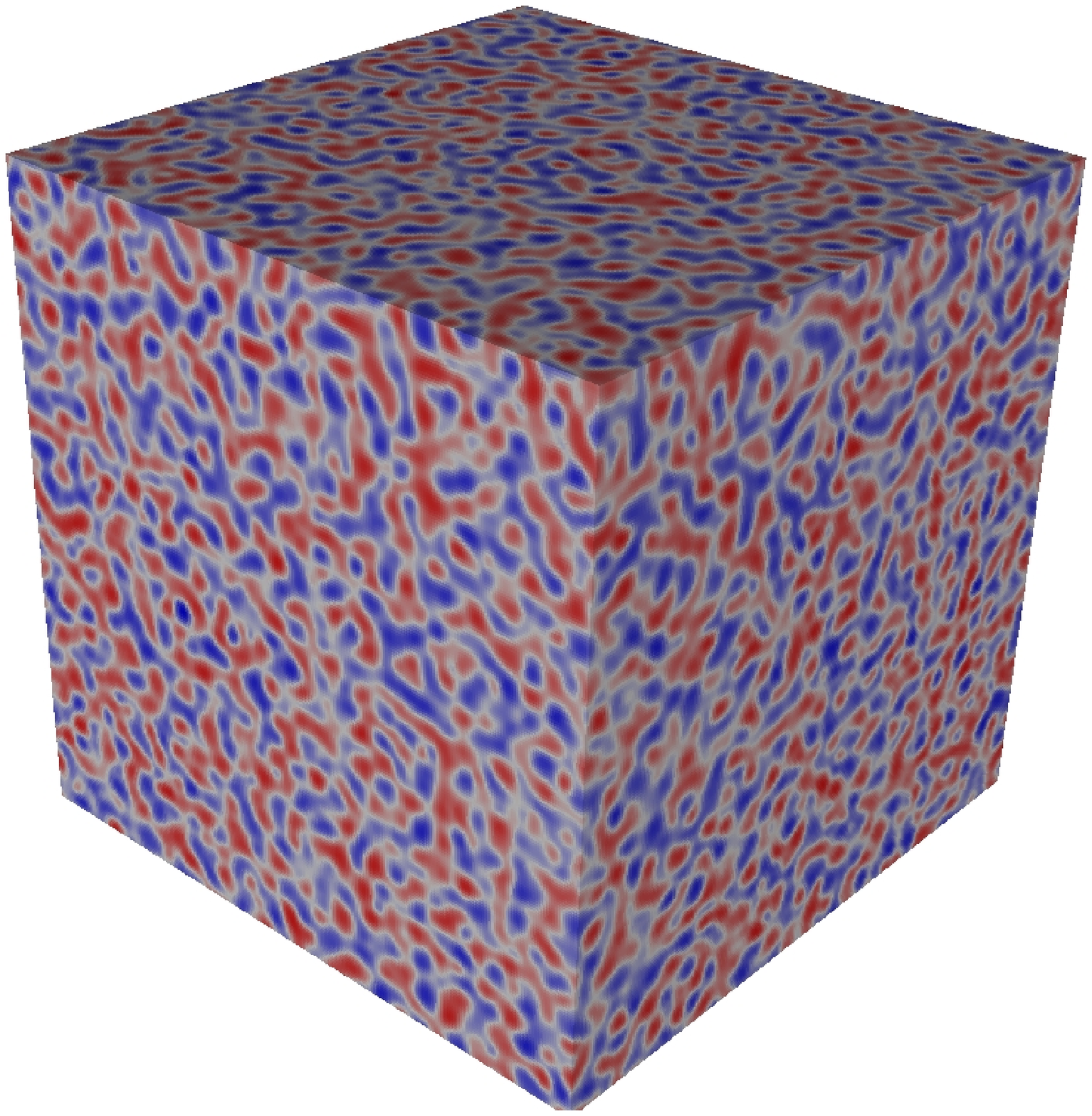}
\includegraphics[width=0.23\linewidth]{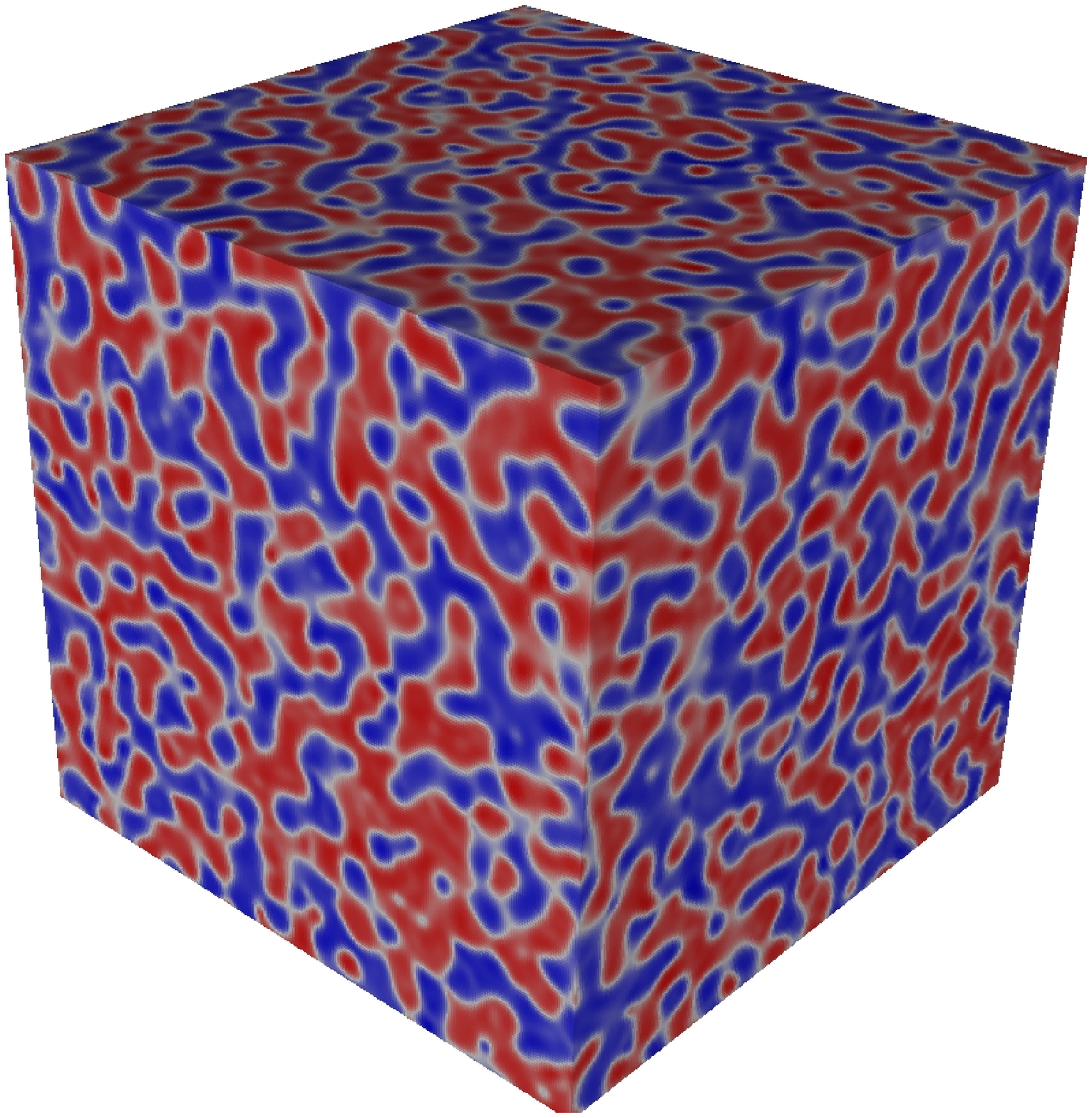}
\includegraphics[width=0.23\linewidth]{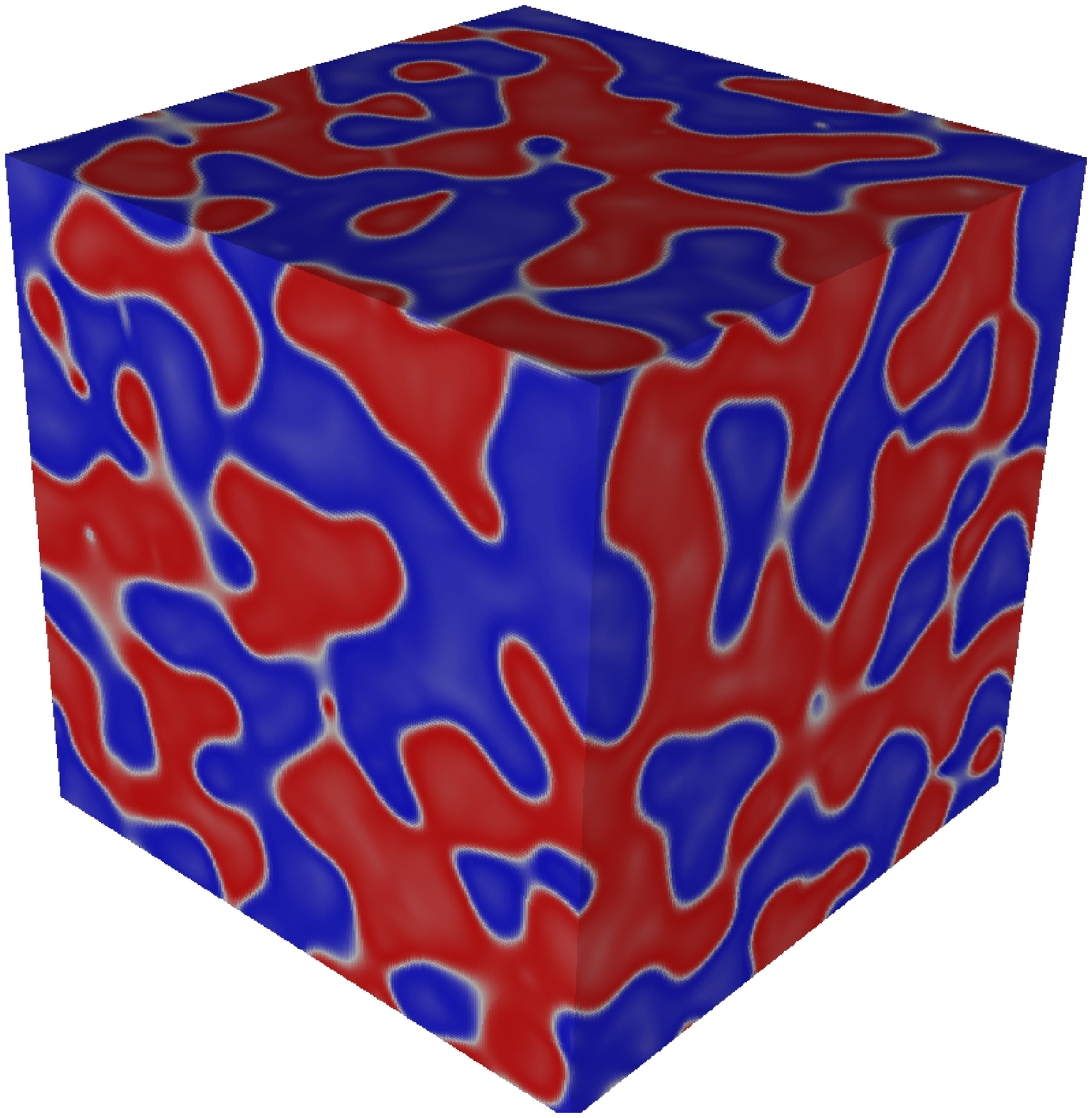}
\includegraphics[width=0.23\linewidth]{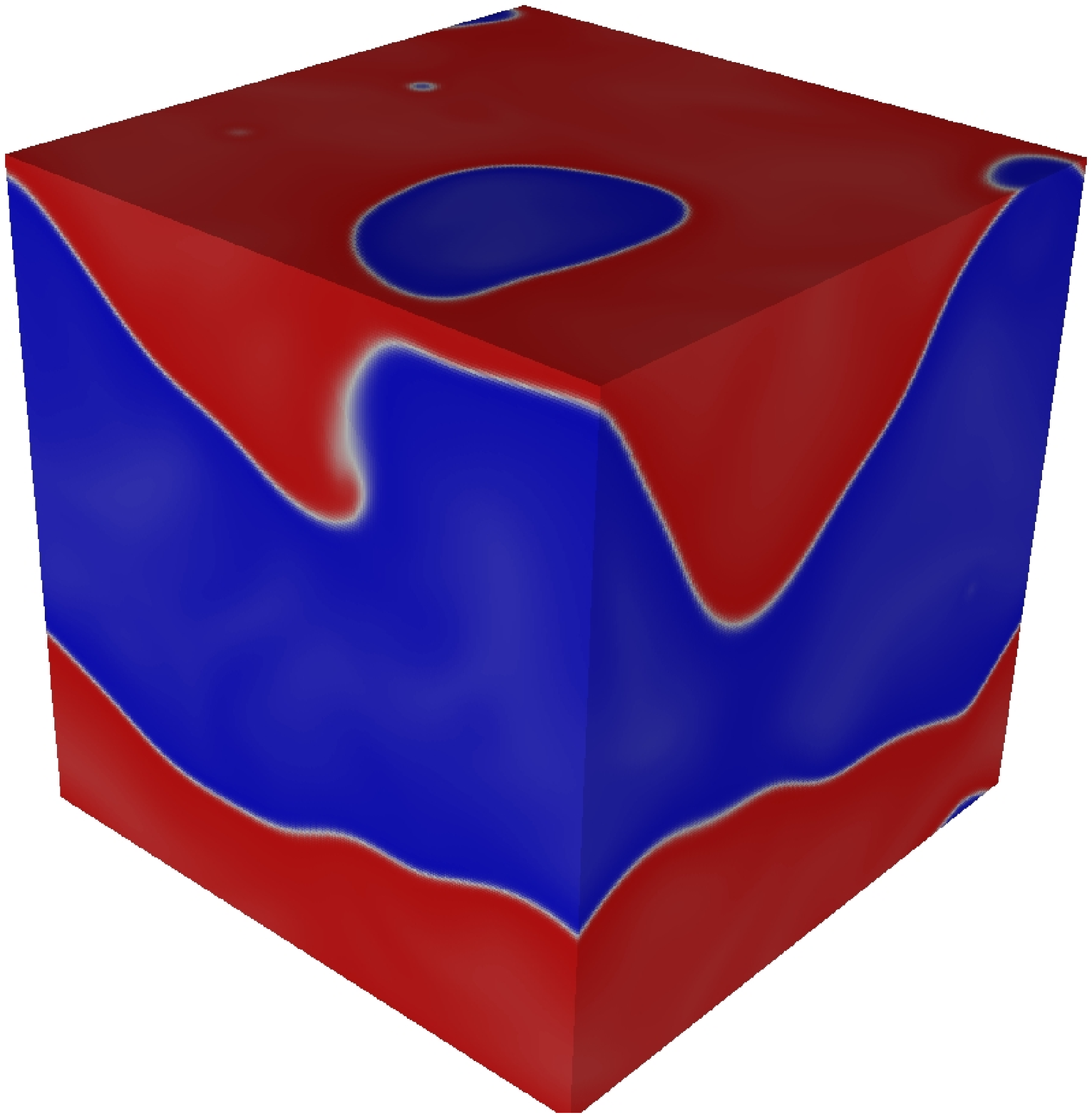}
};
\node[below right](img) at (0,0){
\includegraphics[width=0.23\linewidth]{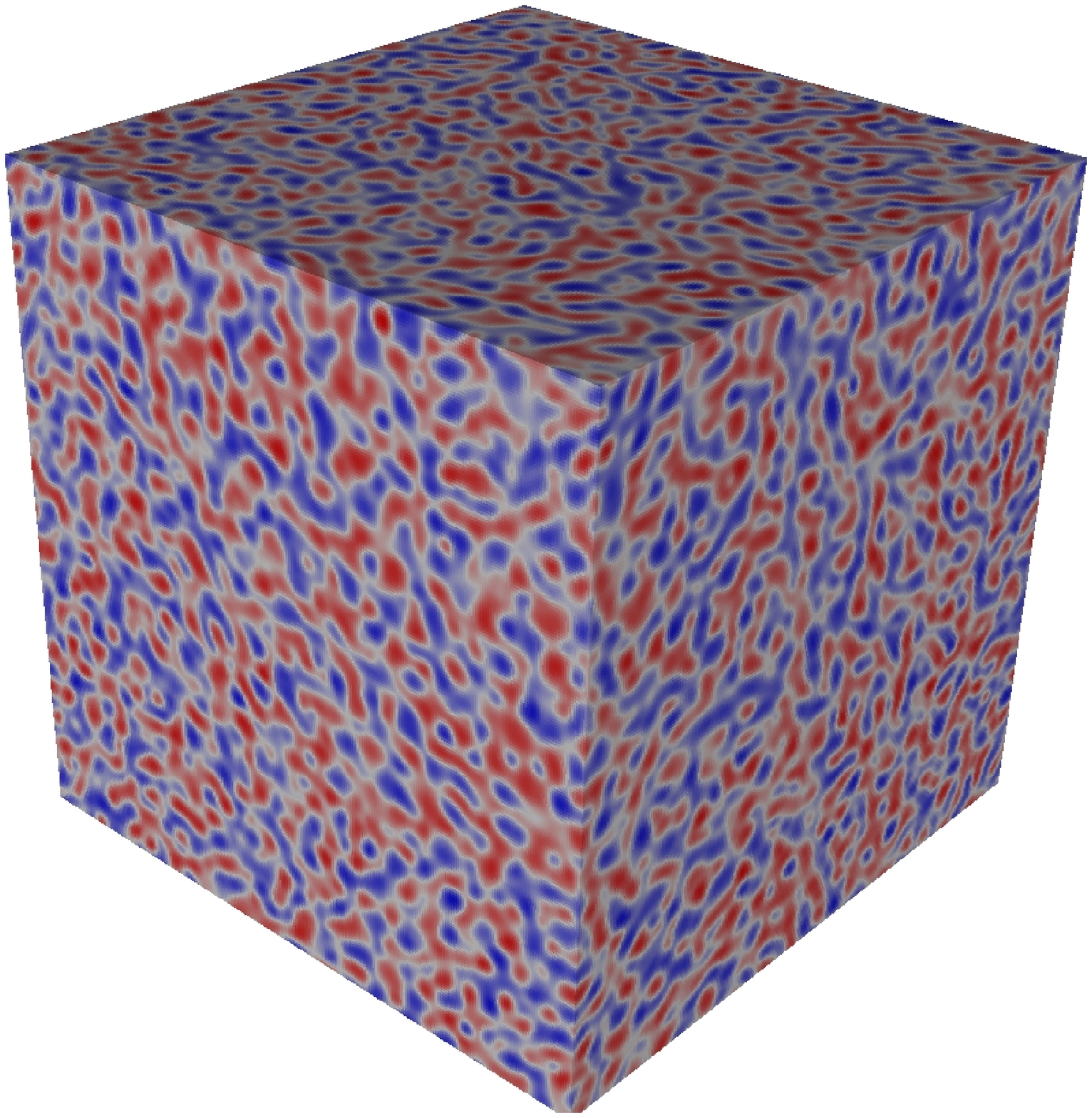}
\includegraphics[width=0.23\linewidth]{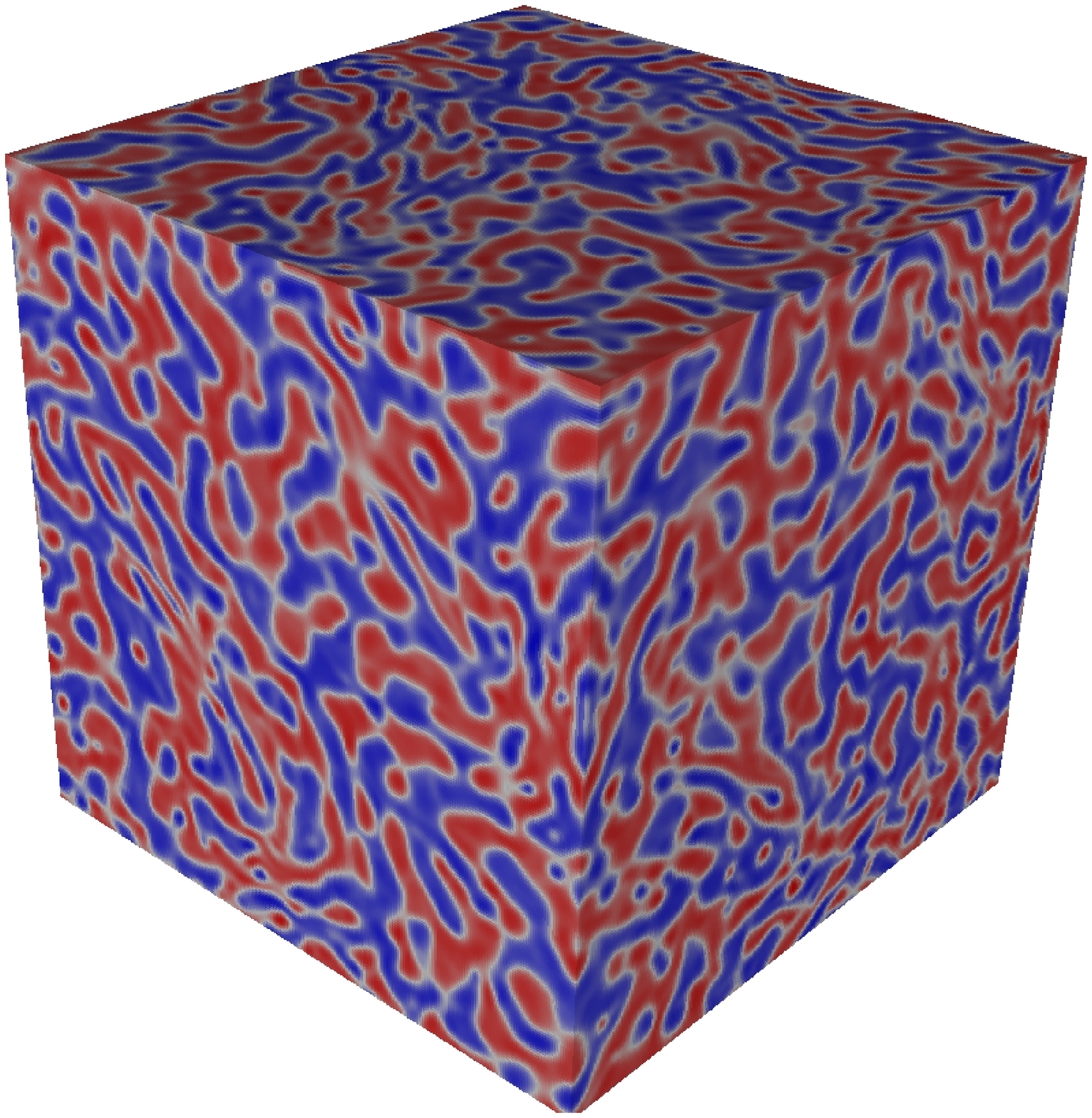}
\includegraphics[width=0.23\linewidth]{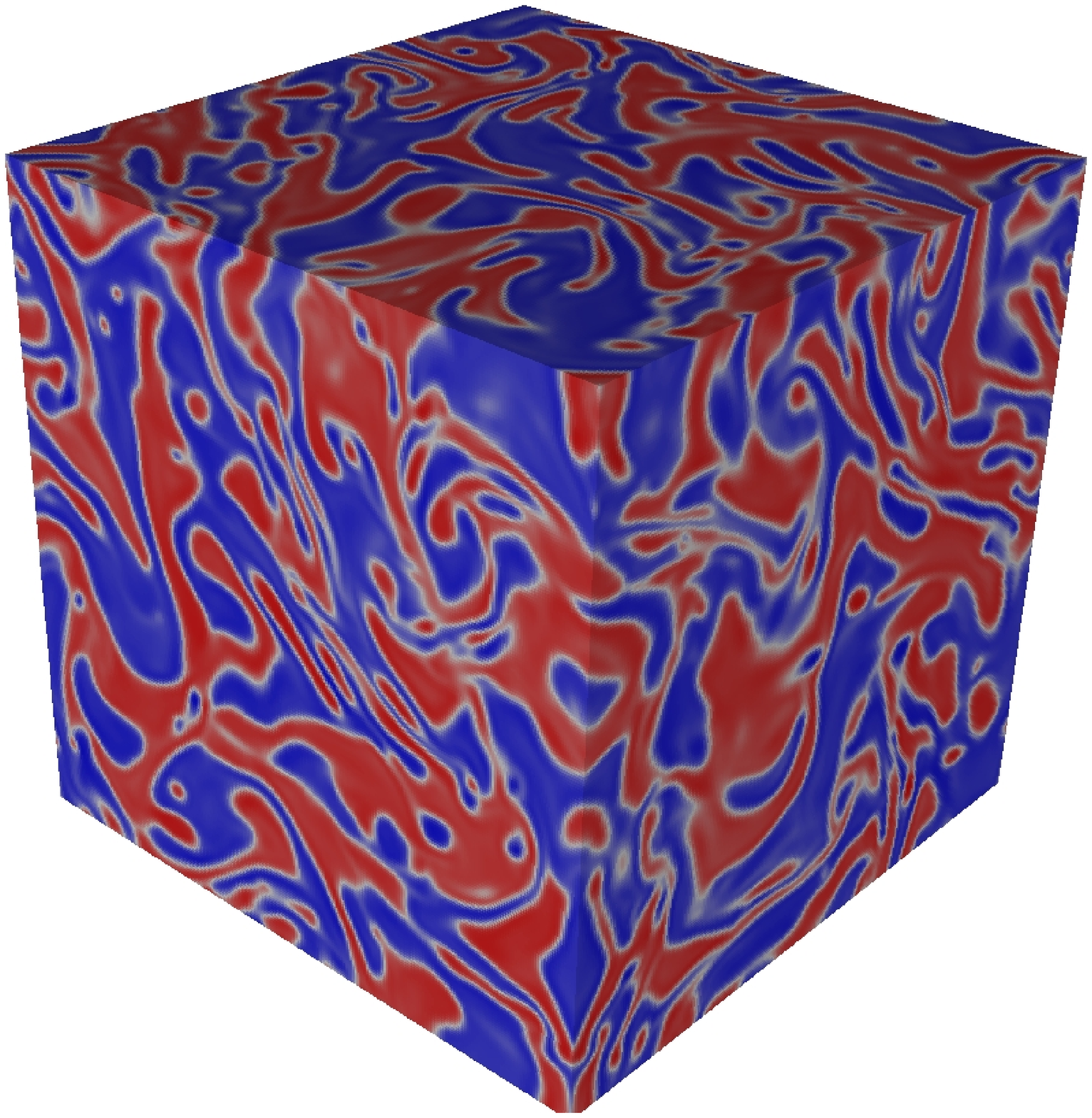}
\includegraphics[width=0.23\linewidth]{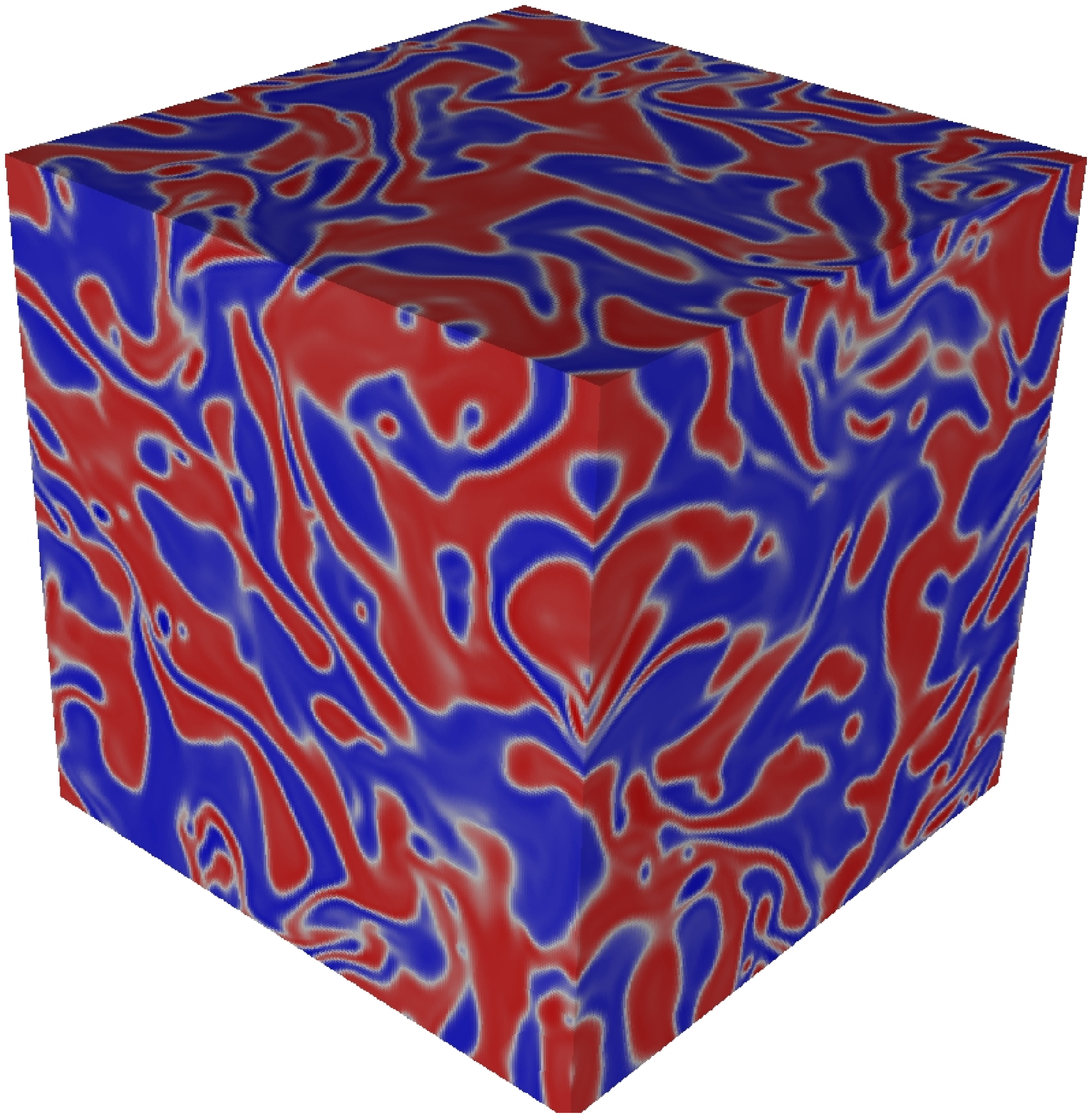}};
\node[below right](img) at (0,-4.2){
\includegraphics[width=0.23\linewidth]{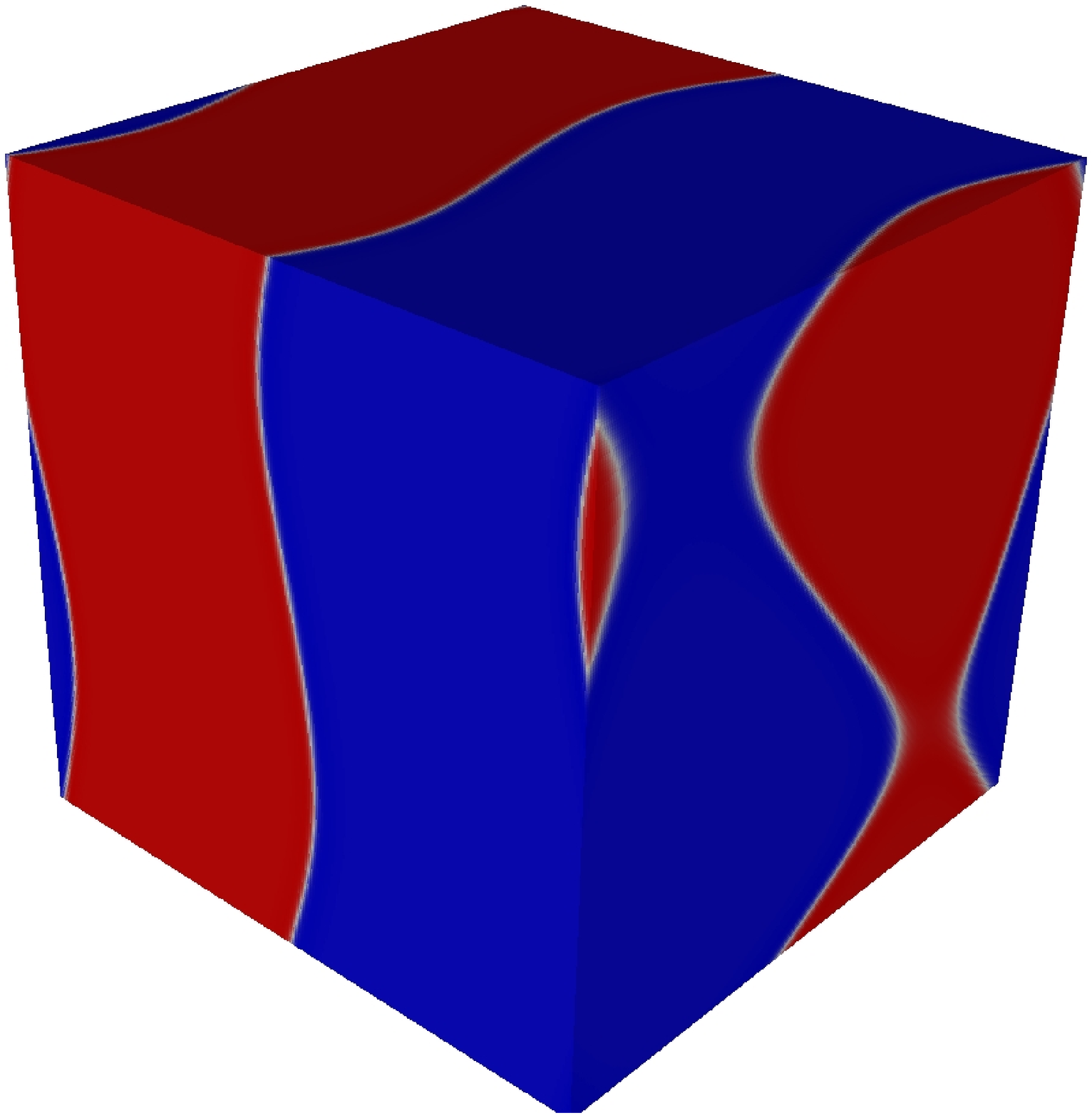}
\includegraphics[width=0.23\linewidth]{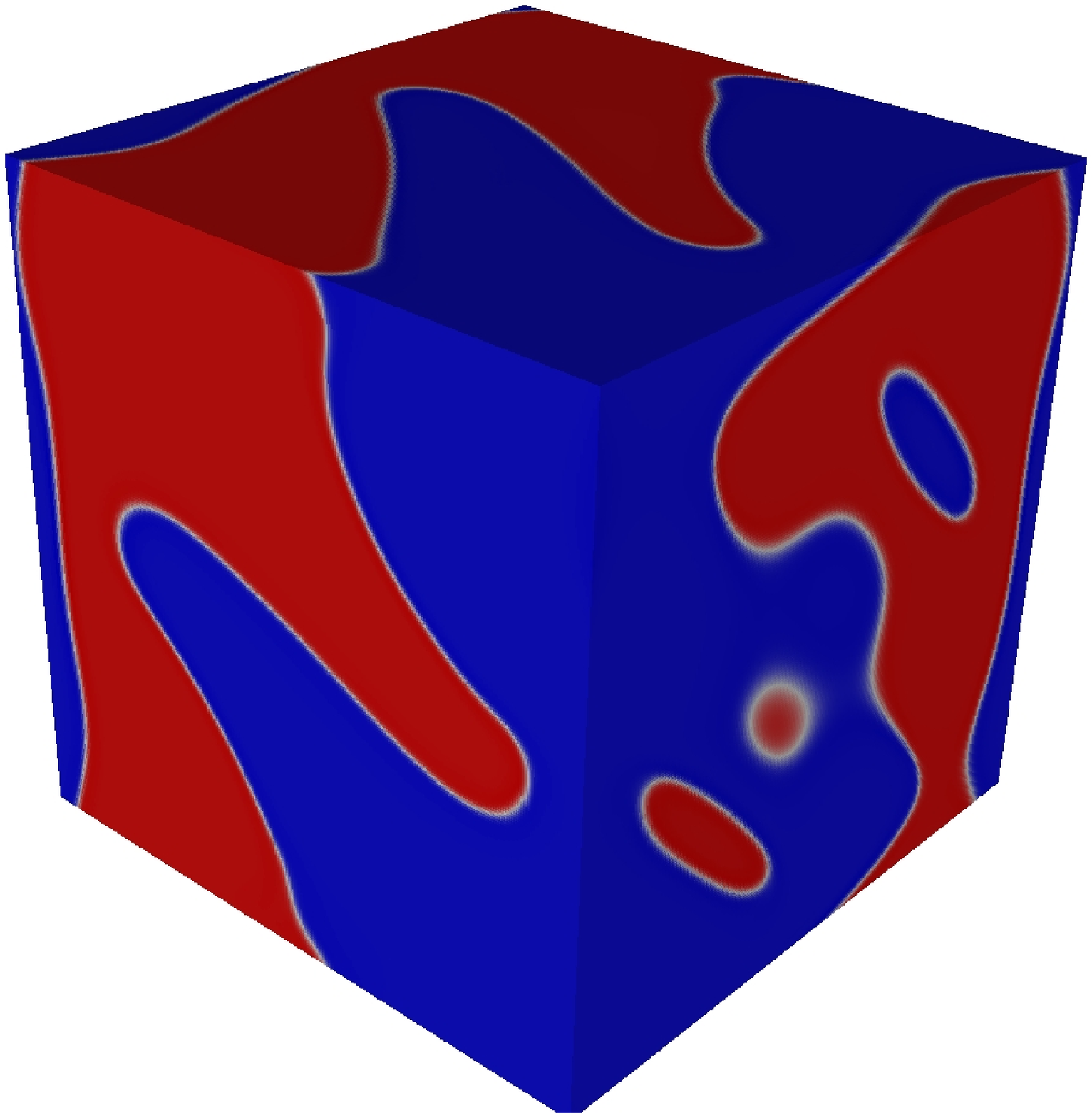}
\includegraphics[width=0.23\linewidth]{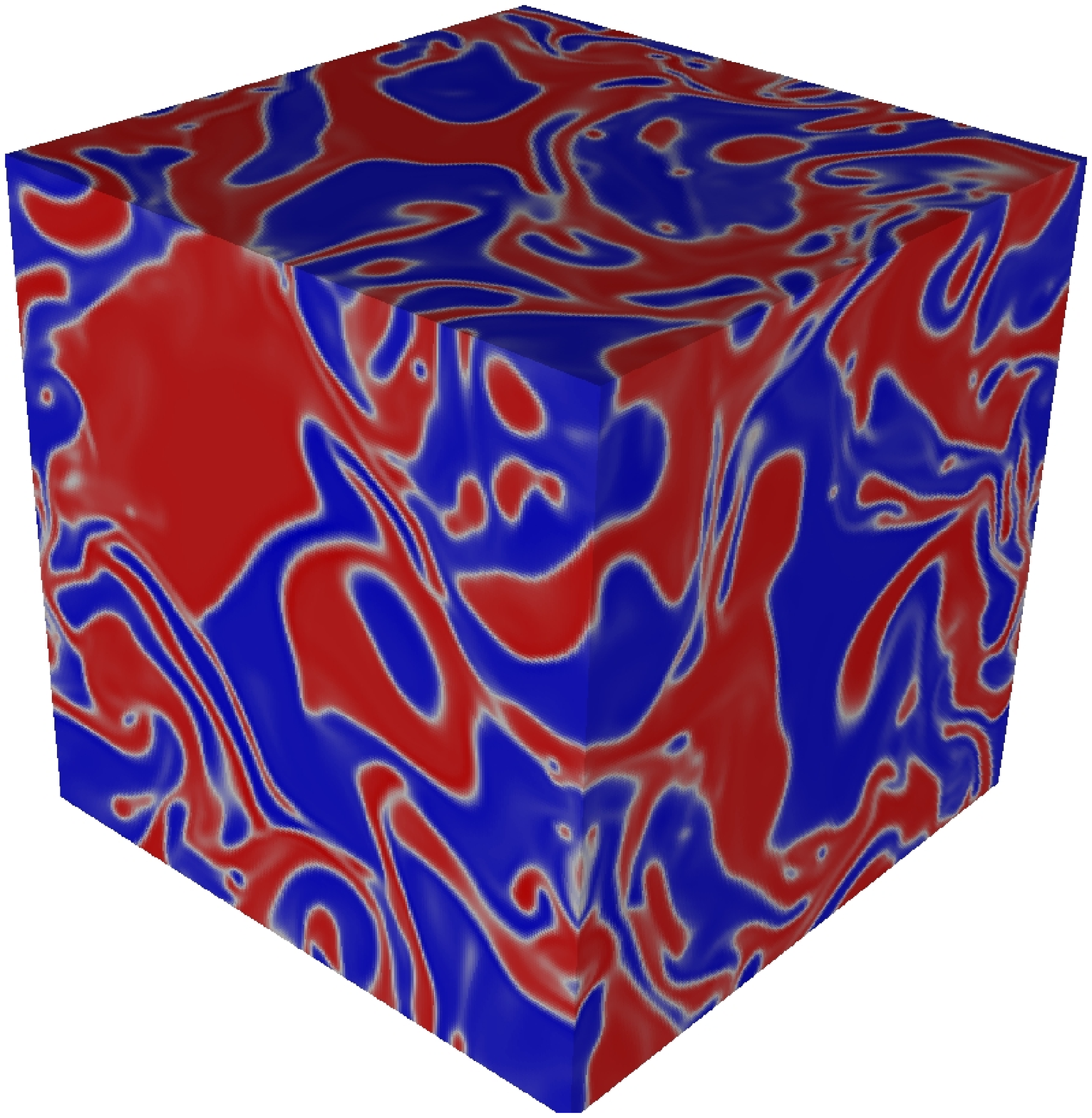}
\includegraphics[width=0.23\linewidth]{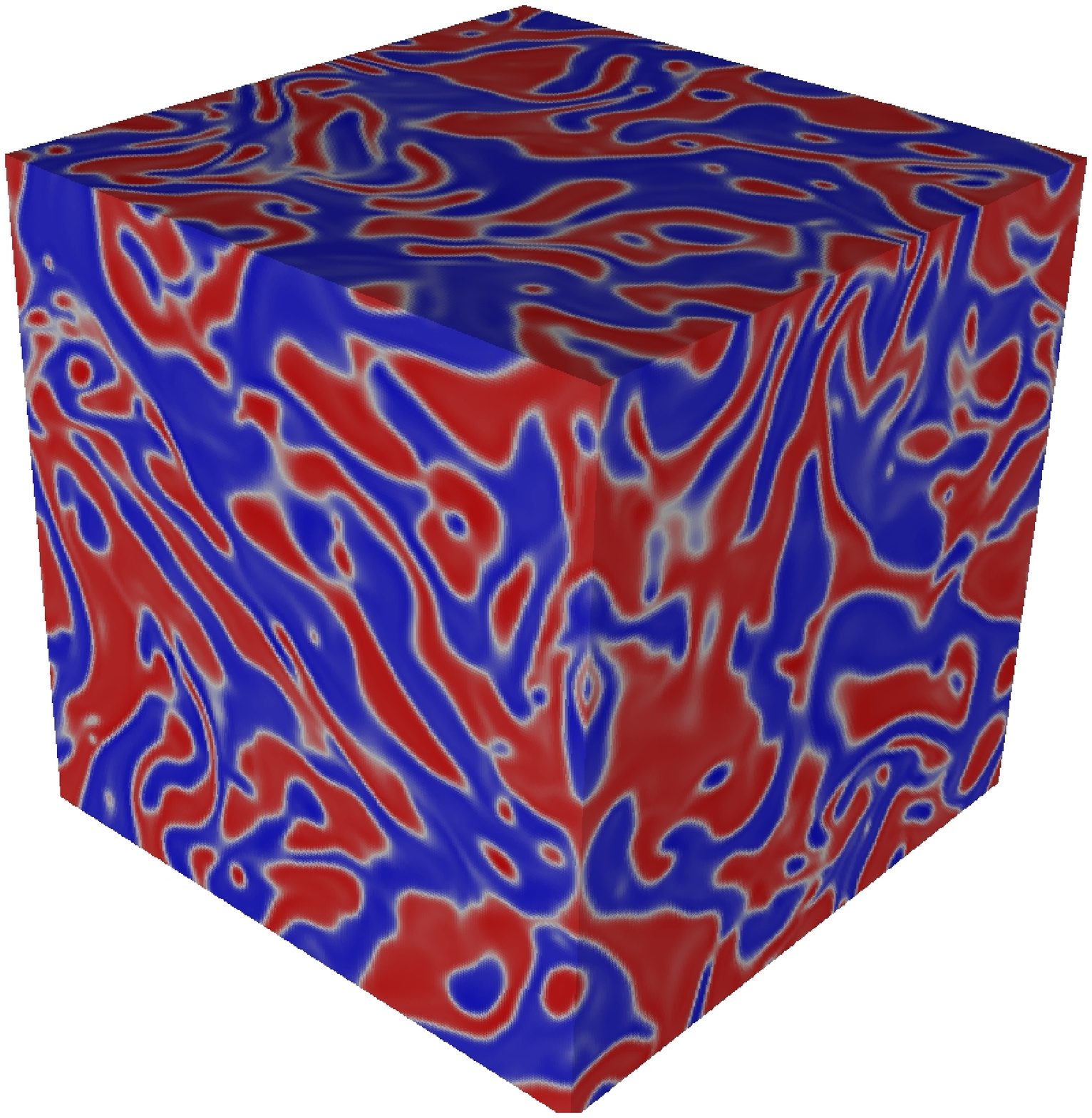}};
\node[below right,rotate=90] at (-0.5,1.0) {\large $No~forcing$};
\node[below right,rotate=90] at (-0.5,-3.4) {\large $Turbulence$};
\node[below right,rotate=90] at (-0.5,-7.4) {\large $Turbulence$};
\node[below right] at (8,-8.3) {\large $time \longrightarrow$};
 \end{tikzpicture}
\caption{\label{fig:ciso} Psuedocolor plots of the concentration
  fields, with the two symmetric fluids indicated in red and blue.
  (Top panel, left-right) Time evolution of the concentration field
  undergoing coarsening process from a initially well-mixed
  state. Notice the formation of ever-larger concentration patches as
  the time evolves. (Middle panel, left-right) Time evolution of the
  concentration field undergoing coarsening process from a well-mixed
  state in presence of turbulence generated by external driving. The
  last two panels indicate that a statistically steady state has
  developed. The coarsening process goes on uninhibited until arrested
  by the turbulence at later times.  (Bottom panel, left-right) Time
  evolution of a very coarse phase-separated mixture in presence of
  turbulence with same intensity as the middle-panel. In this case,
  the domains are broken up until the mixture attains a steady state
  domain size that is the same as the one in the middle-panel. This
  behavior indicates a positive renormalized eddy diffusivity at large
  length scales even though the microscopic diffusion constant is
  negative. The Taylor-microscale Reynolds number for the middle and
  bottom-panel is $Re_{\lambda}=103$ (run ST2,
  Table~\ref{tab:comp}). From the plots it is clear that in case of
  turbulence the coarsening of concentration gets arrested whereas
  coarsening length attains domain size in absence of turbulence. In
  all the panels, the snapshots are taken at times $t=5.0\cdot 10^3,
  1.0\cdot 10^4, 2.5\cdot10^4$, and $1.0\cdot 10^5$.
}
\end{figure*}

In this letter we study the competition between incompressible
turbulence and the coarsening, which leads to a dynamically-active
statistically steady state \cite{rui81}. Turbulence twists, folds, and
breaks interfaces into smaller domains whereas coarsening leads to
domain growth. We show here that turbulence leads to an arrest of the
coarsening length (see Fig.~\ref{fig:ciso}). We present
{state-of-the-art} high-resolution numerical simulations of a
symmetric liquid binary mixture in three-dimensions in presence of
external turbulent forcing. We first verify the presence of the
viscous scaling regime in
the dynamical coarsening in the absence of turbulence, also observed
in earlier simulations of binary mixtures \cite{seg03, pag01}. Next we
add an external forcing at large scales to generate homogeneous and
isotropic turbulence. Our simulations show that the competition
between breakup due to turbulence and coagulation due to spinodal
decomposition leads to coarsening arrest. Our results are quantified
by observing both the evolution of the concentration power spectrum
and the coarsening length scale. We show that the coarsening
length-scale can be estimated in terms of the Hinze criterion for
droplet breakup~\cite{hin55,per12} pointing at a common physics
behind the processes. Finally, we show that the back-reaction of the
binary mixture dynamics on the fluid leads to an alteration of the
energy cascade.

Early experiments~\cite{pin84,ton89} used light scattering to  
investigate the coarsening arrest in high-Schmidt number ($Sc\equiv\nu/D$) 
mixtures where $D$ is the diffusivity of the binary mixture \footnote{Ref. \cite{aro84} use Prandtl number instead of Schmidt. To be consistent with the mixing
  of a scalar in turbulence we use Schmidt number.}. These experiments showed 
that in contrast to the standard coarsening, where a strong light scattering is observed, turbulent stirring induced coarsening arrest exhibits very weak light scattering.  These results
were later understood by invoking the idea of scale-dependent eddy
diffusivity in Ref.\cite{aro84}. There it was argued that the
coarsening would proceed inside the viscous-convective
range~\cite{bat59} where the fluid viscosity is important but the
diffusivity of the binary mixture can be ignored. More recent
numerical simulations in two-dimensions have studied the effect 
of chaotic or random velocity fields on the Cahn-Hilliard equation 
and found that the coarsening is indeed arrested~\cite{lac95, ber01, nar07}. 
Here, the coarsening length is determined by the balance of the advection of the
binary-mixture concentration with gradients in the chemical
potential. In Ref.~\cite{ber05} were reported numerical simulations of
fully coupled Navier-Stokes and Cahn-Hilliard equations (at $Sc=0.1$)
with externally forced turbulence in two-dimensions in the inverse
cascade regime. It was shown that the coarsening length varies as
$u_{rms}^{-0.41}$, where $u_{rms}$ is the root-mean-square velocity.

In this letter, we study coarsening arrest in three-dimensions using 
state of the art numerical simulations.

To simulate binary mixtures we use a two-component Lattice-Boltzmann
method \cite{fri87,Suk07,succ01}. The interaction between the
components are introduced using Shan-Chen algorithm
\cite{sha93,sha94,sha95}.  The Shan-Chen force, that employs density
dependent interactions, is used to introduce non-ideal nature of the
fluid~\citep{sha94}.  
Turbulence is generated by using a large-scale sinusoidal forcing along the three directions. 
All wave-modes whose magnitude is less than $\sqrt{2}$ are active and the phases are chosen 
to be independent  Ornstein-Uhlenbeck processes \cite{per12}.

For all our simulations we used initial conditions with densities
$\rho^{(1)}$ and $\rho^{(2)}$ such that the corresponding initial
order parameter field $\phi\equiv (\rho^{(2)}-\rho^{(1)})/(\rho^{(2)}+\rho^{(1)})$ is a random distribution of $+1$ and $-1$.  For simulations with turbulence the forcing was also switched on at the initial time. We simulate in a cubic domain with periodic boundary conditions on all the sides. Table~\ref{tab:comp} lists the parameters used in our simulations.
\begin{table}[!htdp]
\begin{center}
\begin{tabular}{c c c c c}
\hline
Runs & Domain size & $\rho$ & $Re_{\lambda}$ \\
\hline \hline
S1 & $128^3$ & $2.4$ & $NA$  \\
S2 & $256^3$ & $2.4$ &$NA$  \\
S3 & $256^3$ & $1.1$ & $NA$ \\
\hline
ST1 & $128^3$ & $2.4$ & $35,49$  \\
ST2 & $256^3$ &  $2.4$ & $72,103$  \\
ST3 & $512^3$ & $2.4$ & $103,162,185$  \\
\hline
ST4 & $256^3$ &  $1.1$ & $86$  \\
\hline
NS1 & $256^3$ &  $2.4$ & $103$  \\
NS2 & $512^3$ &  $2.4$ & $162$ \\
\hline
\end{tabular}
\end{center}
\caption{ \label{tab:comp} The parameters of our simulations. Runs
  S1-3 explore spinodal decomposition in absence of an external
  forcing, while ST1-ST4 simulate spinodal decomposition in presence
  of external driving that generates turbulence. For comparing
  turbulent binary mixture with the standard, single-component
  turbulent fluid (i.e. symmetric binary mixture above its critical
  point, without surface tension), we conducted also runs NS1 and NS2.
  For all the runs, the kinematic viscosity $\nu=5\cdot 10^{-3}$. For
  runs S1-3, ST1-3, surface tension $\sigma=1.6\cdot 10^{-3}$, and the
  Schmidt number $Sc=1.47$ whereas, for the run ST4,
  $\sigma=1.7\cdot10^{-3}$, and $Sc=3.72$. The Taylor scale Reynolds 
 number is $Re_{\lambda}\equiv\sqrt{10}E/(\sqrt{\Omega}\nu)$.}
\end{table}

We first investigate how the coarsening proceeds in absence of turbulence in the viscous scaling regime (runs S1-3).   
As a definition of the coarsening length scale $L(t)$ we use  
$L(t) = 2\pi/[k_1(t)]$,  with $k_1(t) = (\sum_k k S_k)/(\sum_k S_k)$, {\rm and}  
$S_k = (\sum_k^{\prime} |\phi_{\bm k}|^2)/(\sum_{k}^{\prime} 1)$.
Here $\phi_{\bm k}$ is the Fourier transform of $\phi$, $S_k$ is the shell-averaged 
concentration spectrum normalized by the corresponding density of states. For the 
sake of brevity we will call $S_k$ the concentration spectrum. Finally, $k=\sqrt{{\bm k}\cdot {\bm k}}$, and 
$\sum^{\prime}$ indicates the summation over all the modes $k \in [k-1/2,k+1/2]$.  Below the 
critical point, phase separation leads to an $L(t)$ that grows with time.

\begin{figure}[!h]
\includegraphics[width=\linewidth]{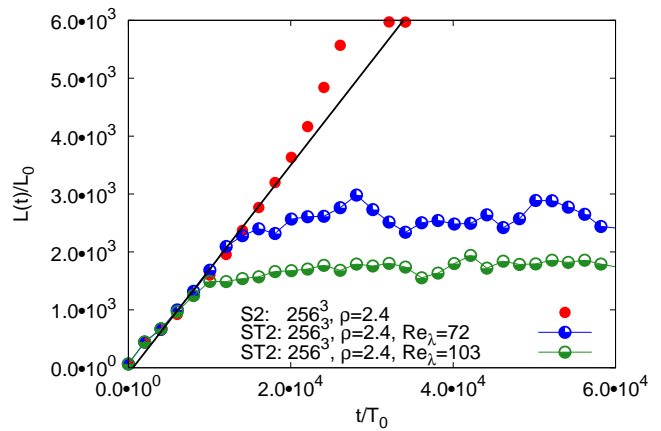}
\caption{\label{fig:turbv} Coarsening arrest for phase separating
  binary mixtures in presence of turbulence. In absence of an external
  turbulent forcing (red circles) the coarsening length keeps on
  growing as $L(t)\sim t$ (black line). Switching on turbulence, the
  coarsening length initially grows undisturbed but then it arrests as
  the system attains a steady state. Note that for fixed $\rho$ and
  surface tension $\sigma$, the saturating coarsening length $L_\infty$ 
 decreases at increasing $Re_{\lambda}$. The time and length scale are non-dimensionalized by the corresponding characteristic length $L_0=\nu^2/(\rho \sigma)$ and time $T_0=\nu^3/(\rho \sigma^2)$.}
\end{figure}

For our runs S1-3, as expected \cite{sig79}, we observe $L(t)\sim t$ [Fig.~\ref{fig:turbv}(red dots)].

We now study the effect of turbulence on coarsening.  We
force large length scales to generate homogeneous, isotropic
turbulence in the velocity field. In what follows, we study the effect
of turbulence in the viscous scaling regime. Note that $Sc=\nu/|D|\sim
{\cal O}(1)$ in our simulations, whereas for most liquid binary mixtures
$Sc\gg1$, which requires the resolution of both inertial and the
viscous-convective scales to be much higher than what is attained in
the present investigation.

Fig.~\ref{fig:turbv} shows how $L(t)$ increases in the presence of
turbulence. Instead of coarsening until $L(t)$ reaches the size of the
simulation domain, turbulence arrests the inverse cascade of
concentration fluctuations, blocks coarsening, and leads to a steady
state length where the domains constantly undergo coalescence and
breakup. The saturating coarsening length $L_\infty$ decreases with
increasing turbulence intensity.

In an earlier study \cite{per12}, we had shown that for asymmetric
binary mixtures the Hinze criterion provides an estimate for the average
droplet diameter undergoing breakup and coalescence in turbulence. We
now show that even for $50\%-50\%$ binary mixtures, the Hinze
criterion gives a good estimate for the coarsening length scale
$L_\infty$ at long times.

According to the prediction of Ref. \cite{hin55}, the maximum droplet
diameter that can be stable to turbulent velocity fluctuations in the
steady state should be given by the Hinze length
\begin{equation}
L_H \approx \left(\frac{\rho}{\sigma}\right)^{-3/5} \epsilon^{-2/5}.
\label{eq:hinze}
\end{equation}
Actually the above equation is also consistent with the predictions  
of Ref.~\cite{cat12}. A general criteria for the coarsening length is given 
by the relation $L(t)=L_0 f(x)$ with $x=t/T_0$ \cite{cat12}.The function $f(x)$ satisfies the two limiting scaling $f(x)\sim x$ for small $x$ and  $f(x)\sim x^{2/3}$ for large $x$. In turbulent flow, the relevant time scale is given by $L/\delta v(L)$ where $\delta v(L)$ is the size of the velocity fluctuation at the scale $L$. Since this time scale is much longer than $t$, the appropriate scaling behavior for $f(x)$ is $x^{2/3}$. Using the inertial scaling we again obtain Eq. \eqref{eq:hinze}. The above argument may not 
apply to shear flows, due to non isotropic contributions and strong dissipation at the boundaries, and in two dimensinal flows where the characteristic time scale is dictated by enstrophy cascade.

\begin{figure}[!h]
\includegraphics[width=\linewidth]{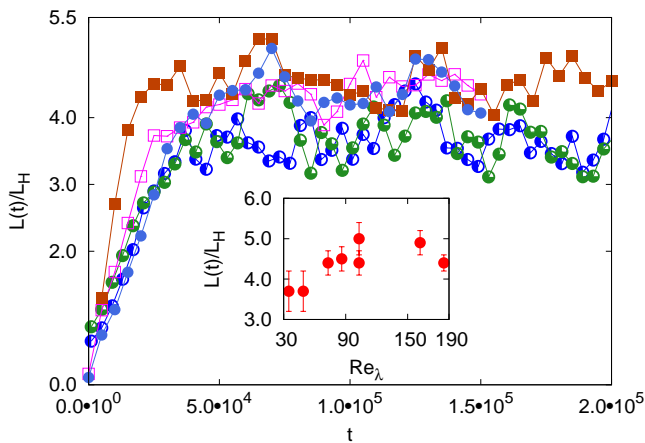}
\caption{\label{fig:turbvh} Growth of the coarsening length scale
  $L(t)$ in the arrested state normalized by the Hinze length $L_H$
  for $Re_{\lambda}=35, 49$ (blue three-quarter-filled circle and
  green half-filled circle) [run ST1], $Re_{\lambda}=72,103$ (purple
  square and brown filled square) [run ST2], and $Re_{\lambda}=86$
  (blue filled circle) [run ST4]. In the inset we plot the average
  value of $L(t)/L_H$, calculated over the time window $t=5\times10^4$ 
  to $2\times10^5$, for different Reynolds numbers
  $Re_{\lambda}$. Within errorbars, $L(t)/L_H\approx 4.4\pm0.5$ is
  found to be a good indicator for the arrested length scale. We
  believe that the smaller value of $L(t)/L_H$ (although within our
  error bars) for the $Re_{\lambda}=35,49$ arises because of the lower
  grid resolution. .}
\end{figure}

In \cite{per12,per11_jpcs} it was shown that in presence of
coagulation-breakup processes the correct quantity to look at is the
average droplet diameter $L_\infty \equiv\langle L(t) \rangle$ in the
statistically stationary state. Therefore we expect that the ratio
$L(t)/L_H$ stay constant in all our simulations. The plot in
Fig.~\ref{fig:turbvh} shows the plot of $L(t)/L_H$ for our runs.

\begin{figure}
\includegraphics[width=\linewidth]{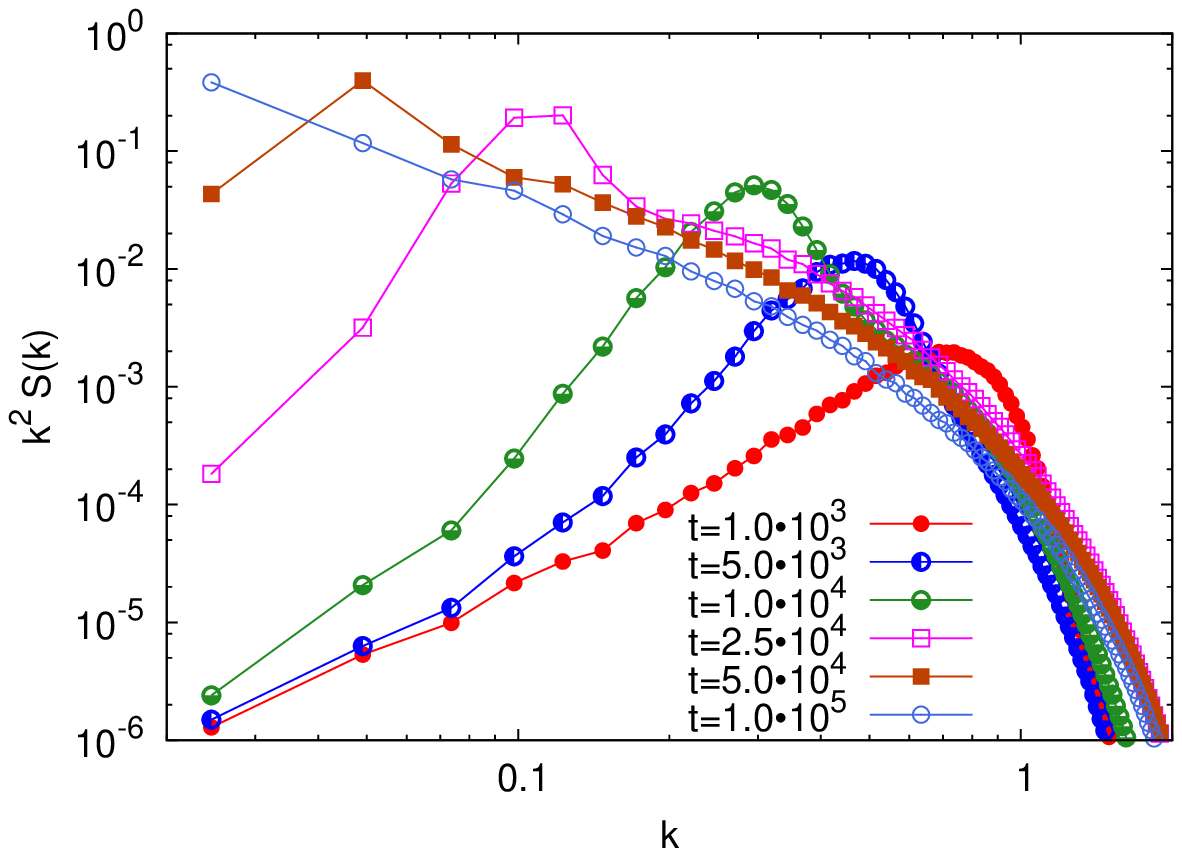}
\includegraphics[width=\linewidth]{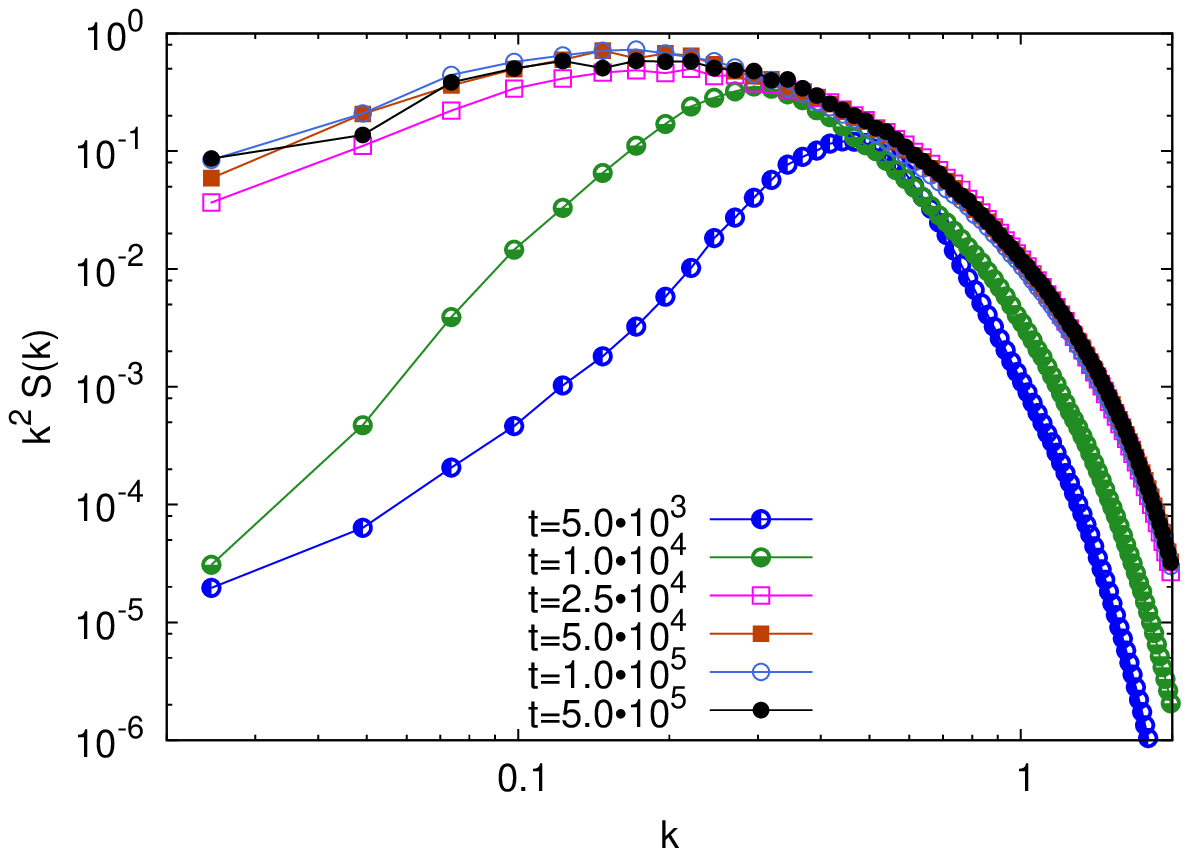}
\caption{\label{fig:ccasc} (Top panel) The inverse cascade of
  concentration spectrum $k^2 S(k)$ for spinodal decomposition at
  times $t=10^3-5\cdot10^5$ ($256^3, \rho=2.4$ [run S2]) without
  turbulence. The Fourier mode associated with the peak of the
  spectrum gives an estimate of the instantaneous coarsening
  length. The peak shifts towards the small wave vectors (large length
  scales) as time progresses.  On the other hand, in presence of
  turbulence (bottom panel, \{$256^3, \rho=2.4$, and $Re_{\lambda}=103$ 
  [run ST2]\}), we do observe an initial inverse cascade of concentration 
  that saturates around $t=5\cdot10^{4}$ indicating a blockage of the inverse
  cascade.}
\end{figure}

The blockage of the energy transfer by turbulence is best understood
in Fourier space. The plots in Figure~\ref{fig:ccasc} compare the
concentration spectrum $k^2 S_k$ at various times $t=10^3, 10^4$, and
$10^5$ in absence ($256^3, \rho=2.4$ [run S2]) and presence ($256^3,
\rho=2.4$, and $Re_{\lambda}=103$ [run ST2]) of turbulence. Without
turbulence we observe a peak in the concentration spectrum at initial
times that moves towards smaller wave vectors until it reaches the
domain size. On the other hand in presence of turbulence, the
concentration fluctuations saturate and the concentration spectrum reaches a
steady state.

\begin{figure}
\includegraphics[width=\linewidth]{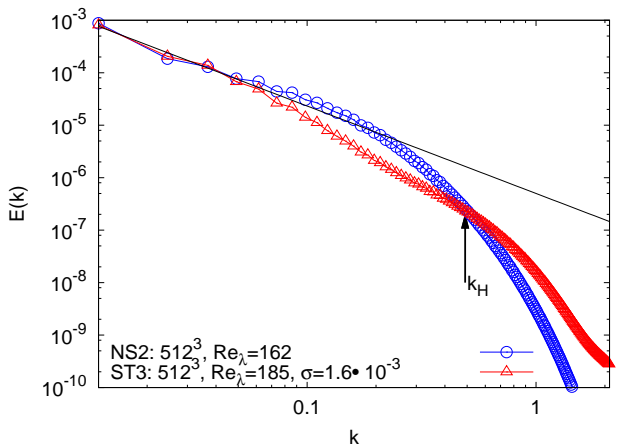}
\caption{\label{fig:vspec} Comparison of the energy spectrum for the spinodal decomposition in presence of turbulence (triangle, run ST3 [$Re_{\lambda}=185$]) with the pure fluid case (circle, run NS2). The black line indicates the Kolmogorov scaling $k^{-5/3}$. We observe that the energy content of the binary mixture is lower than the pure fluid case in the inertial range but in the deep-dissipation range it is higher. We find that the large-$k$ cross-over takes place roughly around the inverse Hinze scale $k_H\equiv 1/L_H$. This cross-over 
was also confirmed by comparing runs NS1 and ST2 (not shown here).}
\end{figure}

The presence of a surface tension should also alter the transfer of
energy in Fourier space. The stretching and folding of the surface by
turbulence takes energy from the fluid. On the other hand in the
regions of weak turbulence local chemical potential will transfer
energy back to the fluid. In Fig.~\ref{fig:vspec} we investigate how a
phase separating binary liquid mixture velocity spectrum compares with
the pure fluid case. We observe that in the inertial range the energy
content of the binary mixture is strongly suppressed in comparison to
the pure fluid case whereas, in the dissipation range the energy
content is higher for the binary mixture. Qualitatively this
phenomenon is similar to the case of dilute polymer solutions in
turbulence where the polymer elasticity is shown to alter the energy
transfer in a similar way \cite{per06,ang05}. However, we note
that the main difference between polymers and binary mixture interface
dynamics is that the size of polymers is smaller than the dissipation
range whereas, in the present study, the droplet size lie in the
inertial range of scales.

We acknowledge the COST Action MP0806 and FOM (Stichting voor
Fundamenteel Onderzoek der Materie) for support.  We acknowledge
computational support from CASPUR (Roma, Italy), from CINECA (Bologna,
Italy), and from JSC (Juelich, Germany). This research was supported
in part by the National Science Foundation under Grant
No. PHY11-25915. Work by DRN was supported by the National Science
Foundation (USA) via Grant DMR 1005289 and through the Harvard
Materials Research Science and Engineering Laboratory, through Grant
DMR 0820484.

\end{document}